\documentclass[pdftex,aps,prl,amsmath,amsfonts,amssymb,superscriptaddress,twocolumn,showpacs]{revtex4}

\usepackage[pdftex]{graphicx}
\usepackage{wasysym}
\usepackage{setspace}
\usepackage{color}

\newcommand{\figref}[1]{Fig.~\ref{fig:#1}}
\newcommand{\Figref}[1]{Figure~\ref{fig:#1}}

\renewcommand{\eqref}[1]{Eq.~(\ref{eq:#1})}

\def\a{s}
\def\b{s}
\newcommand{\add}[1]{\if\a\b{{\color{red} #1}}\else{#1}\fi}
\newcommand{\comm}[1]{\if\a\b{{\color{blue}\{\small \sc #1\}}}\else{}\fi}
\newcommand{\del}[1]{{\if\a\b{{\color{magenta}[[#1]]}}\else{}\fi}}

\begin{document}

\title{Non-touching nanoparticle diclusters bound by repulsive and attractive Casimir forces}
\author{Alejandro W. Rodriguez}
\affiliation{Department of Physics,
Massachusetts Institute of Technology, Cambridge, MA 02139}
\author{Alexander P. McCauley}
\affiliation{Department of Physics,
Massachusetts Institute of Technology, Cambridge, MA 02139}
\author{David Woolf}
\affiliation{Department of Applied Physics,
Harvard University, Cambridge, MA 02139}
\author{Federico Capasso}
\affiliation{Department of Applied Physics,
Harvard University, Cambridge, MA 02139}
\author{J. D. Joannopoulos}
\affiliation{Department of Physics,
Massachusetts Institute of Technology, Cambridge, MA 02139}
\author{Steven G. Johnson}
\affiliation{Department of Mathematics,
Massachusetts Institute of Technology, Cambridge, MA 02139}

\begin{abstract}
  We present a scheme for obtaining stable Casimir suspension of
  dielectric nontouching objects immersed in a fluid, validated here
  in various geometries consisting of ethanol-separated dielectric
  spheres and semi-infinite slabs. Stability is induced by the
  dispersion properties of real dielectric (monolithic) materials. A
  consequence of this effect is the possibility of stable
  configurations (clusters) of compact objects, which we illustrate
  via a ``molecular'' two-sphere dicluster geometry consiting of two
  bound spheres levitated above a gold slab. Our calculations also
  reveal a strong interplay between material and geometric dispersion,
  and this is exemplified by the qualitatively different stability
  behavior observed in planar versus spherical geometries.
\end{abstract}

\maketitle

Electromagnetic fluctuations are the source of a macroscopic force
between otherwise neutral objects known as the Casimir
effect~\cite{casimir, Lifshitz80, milton04}. In most geometries
involving vacuum-separated metallic or dielectric objects (with a
separating plane), the force is attractive and decaying as a function
of object separation, and may contribute to ``stiction'' in
microelectromechanical systems~\cite{hochan1}. A repulsive interaction
would be desirable to combat stiction as well as for frictionless
suspension and other applications.  Repulsive Casimir forces occur in
a variety of settings, including theoretical magnetic
materials~\cite{Boyer74, Kenneth02, Rosa08}, fluid-separated
dielectrics~\cite{Dzyaloshinskii61, Munday09}, interleaved metallic
geometries~\cite{RodriguezJo08:PRA}, and have also been suggested for
composite metamaterials~\cite{Zhao09} (although repulsion with
physical metamaterials has not yet been clearly demonstrated, as
discussed below). In this letter, we demonstrate stable Casimir
suspension of realistic dielectric/metallic objects immersed in a
fluid. Unlike previous work~\cite{Rodriguez08:PRL, RahiZa09:arxiv},
this suspension does not involve one object enclosing another, but
instead occurs between objects on opposite sides of an imaginary
separating plane.  This effect is a consequence of material
dispersion, and is here validated in various experimentally accessible
geometries consisting of ethanol-separated dielectric spheres and
semi-infinite slabs. Furthermore, we show the possibility of achieving
Casimir ``molecular'' clusters in which objects can form stable
non-touching configurations in space---this is illustrated in a
``diatomic'' or ``dicluster'' geometry involving two dielectric
(silicon/teflon) spheres of different radii bound into a non-touching
pair and levitated above a gold slab.  Finally, our calculations
reveal interesting effects related to the interplay of geometric and
material dispersion, in which stability responds to finite size in a
way that is qualitatively different in planar or spherical geometries.

\begin{figure}[t]
\includegraphics[width=0.9\columnwidth]{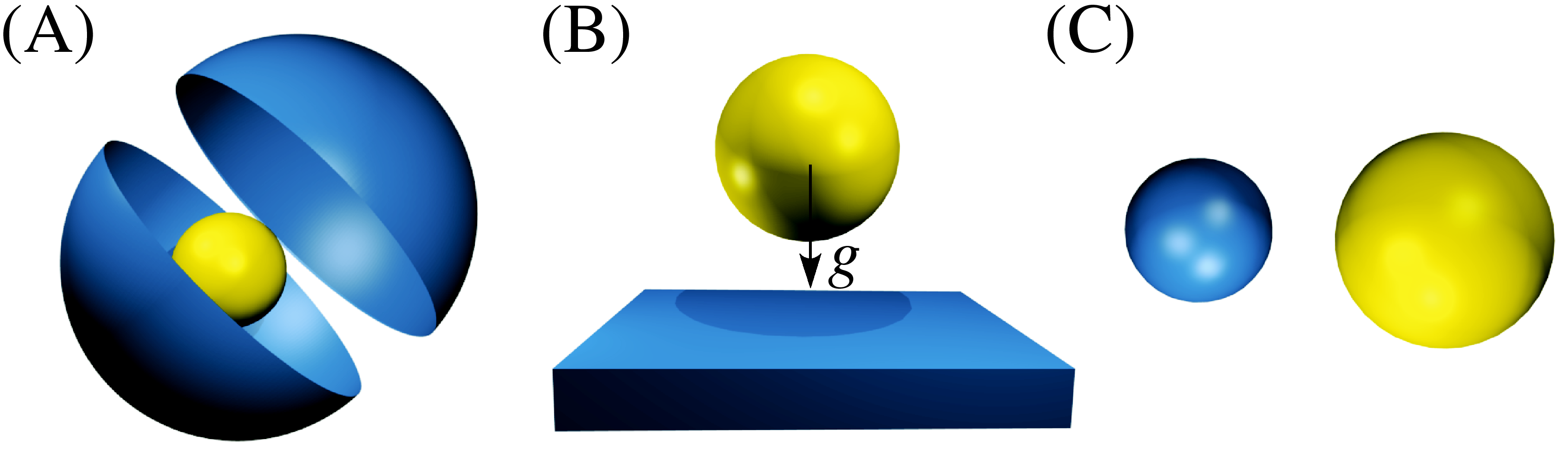}
\caption{Schematic schemes for stable suspension of
  fluid-separated objects, involving: (a) enclosed geometries; (b)
  gravity countering Casimir repulsion; and (c) material dispersion
  producing repulsive \emph{and} attractive Casimir forces (here).}
\label{fig:schem}
\end{figure}

Casimir stability has been previously studied in at least four
different contexts: first, in geometries involving mutually enclosed
fluid-separated objects, in which the inner object is repelled by the
outer object~\cite{Rodriguez08:PRL, RahiZa09:arxiv}; second, a
slab--sphere geometry in which fluid-induced repulsion counteracts the
force of gravity~\cite{McCauleyRo09:PRA}; third, interleaved
structures like the zipper geometry of~\cite{RodriguezJo08:PRA}, in
which the surfaces of two complicated objects interleave so that their
mutual attraction acts to separate the objects; and fourth,
metamaterial proposals~\cite{Zhao09} that currently have no clear
physical realization.  The first two approaches (involving fluids) are
illustrated by the schematics in \figref{schem}(a--b). While all of
these examples clearly demonstrate the possibility of Casimir
stability, they leave something to be desired: they are limited to
enclosed/complex geometries or require that stability lie only along a
single direction (e.g. direction of gravity). It has been proposed
that vacuum-separated chiral metamaterials may exhibit repulsive
interactions and stable repulsive/attractive
transitions~\cite{Zhao09}, but the predicted repulsive forces arise
only for small separations where the metamaterial approximation cannot
be trusted, and recent exact calculations for the proposed structures
indicate that they appear to be
attractive~\cite{McCauley10}. Moreover, recent theoretical work has
shown that vacuum-separated objects can never form stable
configurations~\cite{Rahi10:PRL}. A less constrained and previously
unexplored form of stability is one involving compact objects on
either side of an imaginary separating plane, as illustrated in
\figref{schem}(c) for two spheres: in this case (involving fluids), we
will show that the objects form stable configurations that are
independent of external forces, and seem more accessible to
experiment, opening up new possibilities for the creation of
multi-body clusters based on the Casimir force.

The Casimir force between two dielectric objects embedded in a fluid
can become repulsive if their dielectric permittivities satisfy:
\begin{equation}
  \varepsilon_1(i\xi) < \varepsilon_{\mathrm{fluid}}(i\xi) <
  \varepsilon_2(i\xi),
\label{eq:eps-stab}
\end{equation}
over a sufficiently wide range of imaginary frequencies
$\xi$~\cite{Dzyaloshinskii61}.  The possibility of stable separations
arises if the force transitions from repulsive at small separations
(conceptually dominated by large-$\xi$ contributions) to attractive at
large separations (conceptually dominated by small-$\xi$
contributions).  A criterion for obtaining stability is therefore that
\eqref{eps-stab} be violated at small $\xi$, and satisfied for $\xi >
\xi_c$, with the transition occurring at some critical $\xi_c \sim
2\pi c/\lambda_c$ roughly related to the lengthscale $\lambda_c$ at
which the repulsive/attractive transition occurs.  This criterion is
only heuristic, but helps guide our intuition. The real system is more
complicated, as we shall see, because the sign of the force also
depends on many other factors such as the relative strength of the
contributions of different frequencies (related to the strength of the
$\varepsilon$ contrast) as well as on finite-size effects.

\begin{figure}[t]
\includegraphics[width=0.95\columnwidth]{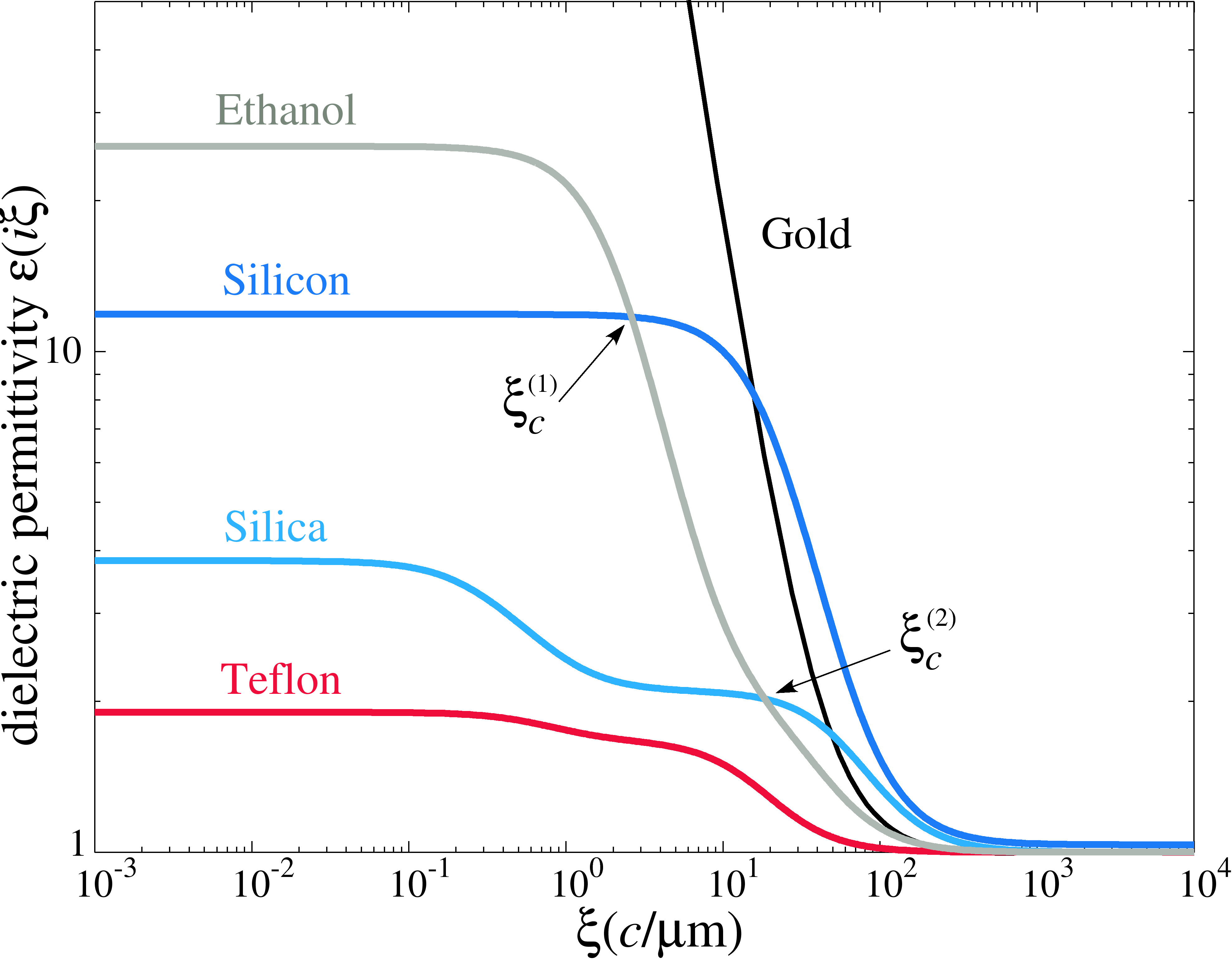}
\includegraphics[width=1.0\columnwidth]{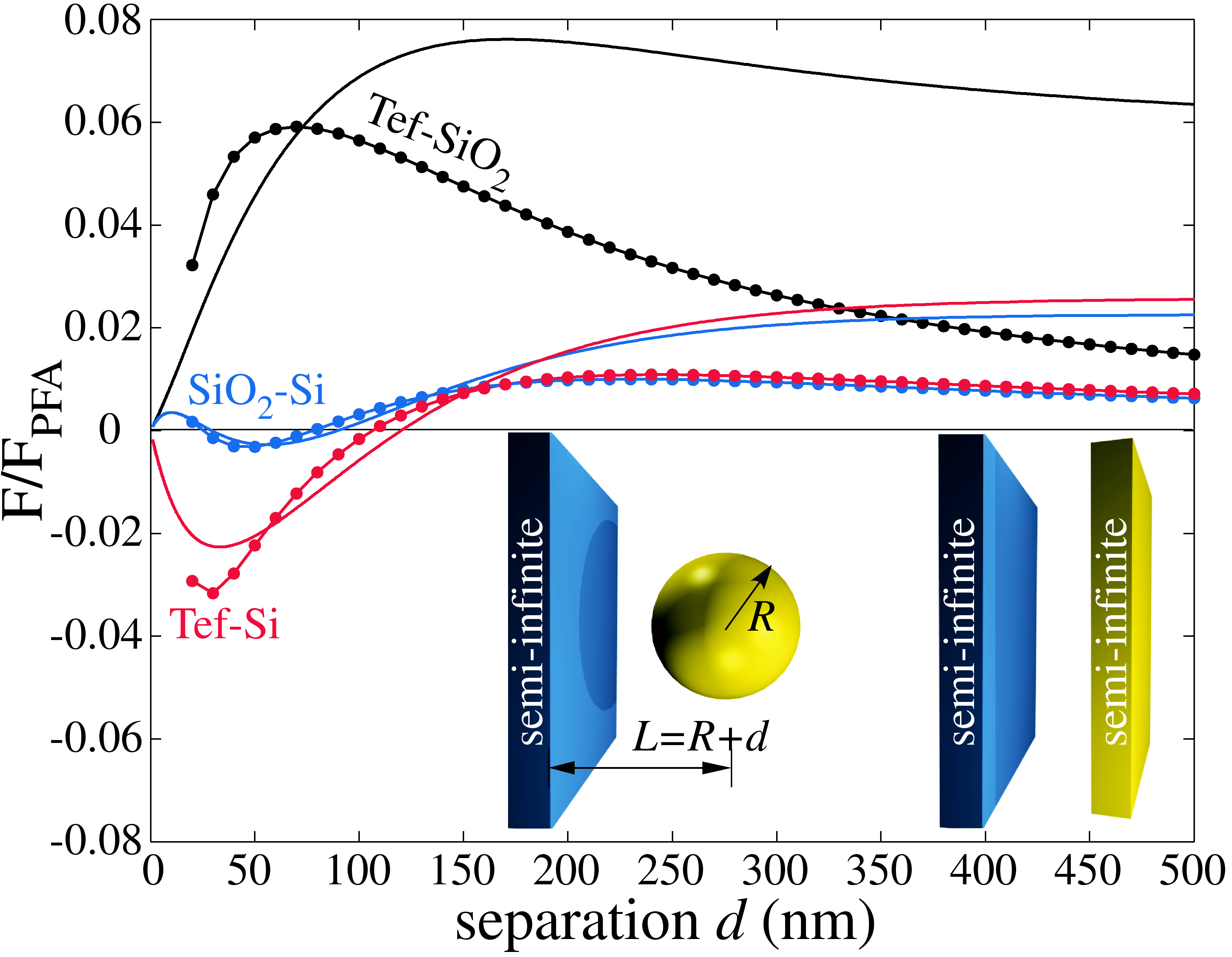}
\caption{(top): Plot of the dielectric permittivity
  $\varepsilon(i\xi)$ of various materials evaluated at imaginary
  frequency $\xi$ (in units of $2\pi c/\mu$m). (Bottom:) Casimir force
  between a semi-infinite slab and a sphere (dots) or semi-finite slab
  (lines), normalized by the corresponding perfect-metal PFA force
  $F_{\mathrm{PFA}} = \hbar c \pi^3 / 720 d^3$ (slab-sphere) or
  $F_{\mathrm{PFA}} = \hbar c \pi^2 / 240 d^4$ (slab-slab). The force
  is plotted for various material configurations, explained in the
  text.}
\label{fig:eps}
\end{figure}

After considering a number of possibilities, we have identified
several material combinations that satisfy \eqref{eps-stab} for large
$\xi$ (small separations). \Figref{eps}(top) plots the dielectric
permittivity $\varepsilon(i\xi)$ of various materials (Si, teflon,
SiO$_2$, and ethanol) satisfying \eqref{eps-stab} over some large
range of $\xi$. In order to establish the existence of stable
separations, we first compute the force between semi-infinite slabs
separated by ethanol, using the Lifshitz
formula~\cite{Lifshitz80}. \Figref{eps}(bottom) plots the Casimir
force between semi-infinite slabs for different material arrangements,
normalized by the corresponding force between perfectlly-metallic
slabs. Our results show that both teflon--Si (yellow) and SiO$_2$--Si
exhibit stable equilibria at $d_c \approx 120$nm and $d^{(s)}_{c}
\approx 90$nm, respectively. SiO$_2$--Si exhibits a finite region of
stability coming from the existence of an \emph{unstable} equilibrium
at a smaller $d^{(u)}_c \approx 29$nm, a consequence of its two
dielectric crossings $\xi^{(1)}_c \approx 2.6 \, 2\pi c/\mu$m and
$\xi^{(2)}_c \approx 26.4 \, 2\pi c/\mu$m labelled in
\figref{eps}(top).  As mentioned above, this prescription for
obtaining stability is only heuristic: for example, the force between
teflon--SiO$_2$ slabs is always attractive, even though their
permittivities satisfy \eqref{eps-stab} at large $\xi$. In that case,
while there is a crossing of the form of \eqref{eps-stab} at
$\xi^{(2)}_c$, the repulsive contributions coming from $\xi >
\xi^{(2)}_c$ are overwhelmed by the attractive contributions coming
from $\xi < \xi^{(2)}_c$, since SiO$_2$ and teflon become transparent
at relatively small $\xi$. Thus, the repulsive region of the frequency
spectrum merely \emph{reduces} the attractive force between the
objects at small separations.

The existence of stable separations for semi-infinite slabs is
promising, especially because this occurs in the 100nm range where
Casimir forces are easily observable, leading us to investigate
whether similar phenomena occur for finite-size objects:
finite-thickness slabs, slab--sphere, and sphere--sphere
configurations.  For slab--sphere and sphere--sphere geometries, rapid
exact calculations are performed using the spherical-harmonic
scattering formulation of~\cite{Lambrecht06, Rahi09:PRD}, which
converges exponentially fast (we only required spherical harmonics up
to order $\ell = 20$ to obtain better than 1\% accuracy at relevant
separations).  Intuitively, one might expect the finite size or
thickness of an object to suppress the contributions from small $\xi$
(large ``wavelengths''), and therefore to change (or eliminate) the
separation at which the repulsive/attractive transition occurs.  Here,
where attraction comes from small-$\xi$ contributions, one might
expect the finite size to decrease the attractive contributions and
therefore increase the equilibrium separation $d_c$.

\Figref{eps}(bottom) plots the force between a sphere of radius
$R=200$nm and a semi-infinite slab for different material
arrangements, normalized by the corresponding perfect-metal
proximity-force approximation (PFA) force. (When referring to a
geometry consisting of a semi-infinite $\alpha$ and finite $\beta$
object, we shall denote the combination by $\alpha$--$\beta$. For
example, when referring to the force between a semi-infinite teflon
slab and a finite SiO$_2$ sphere, we will use the notation
Tef--SiO$_2$.) The results in this slab--sphere geometry look
qualitatively similar to those in the slab--slab structure. In
particular, both Si--Tef and Si--SiO$_2$ exhibit stable equilibria at
$d_c \approx 105$nm and $d_c \approx 78$nm, respectively, roughly
$15$nm smaller than the $d_c$ in the slab-slab case.

\begin{figure}[t]
\includegraphics[width=1.0\columnwidth]{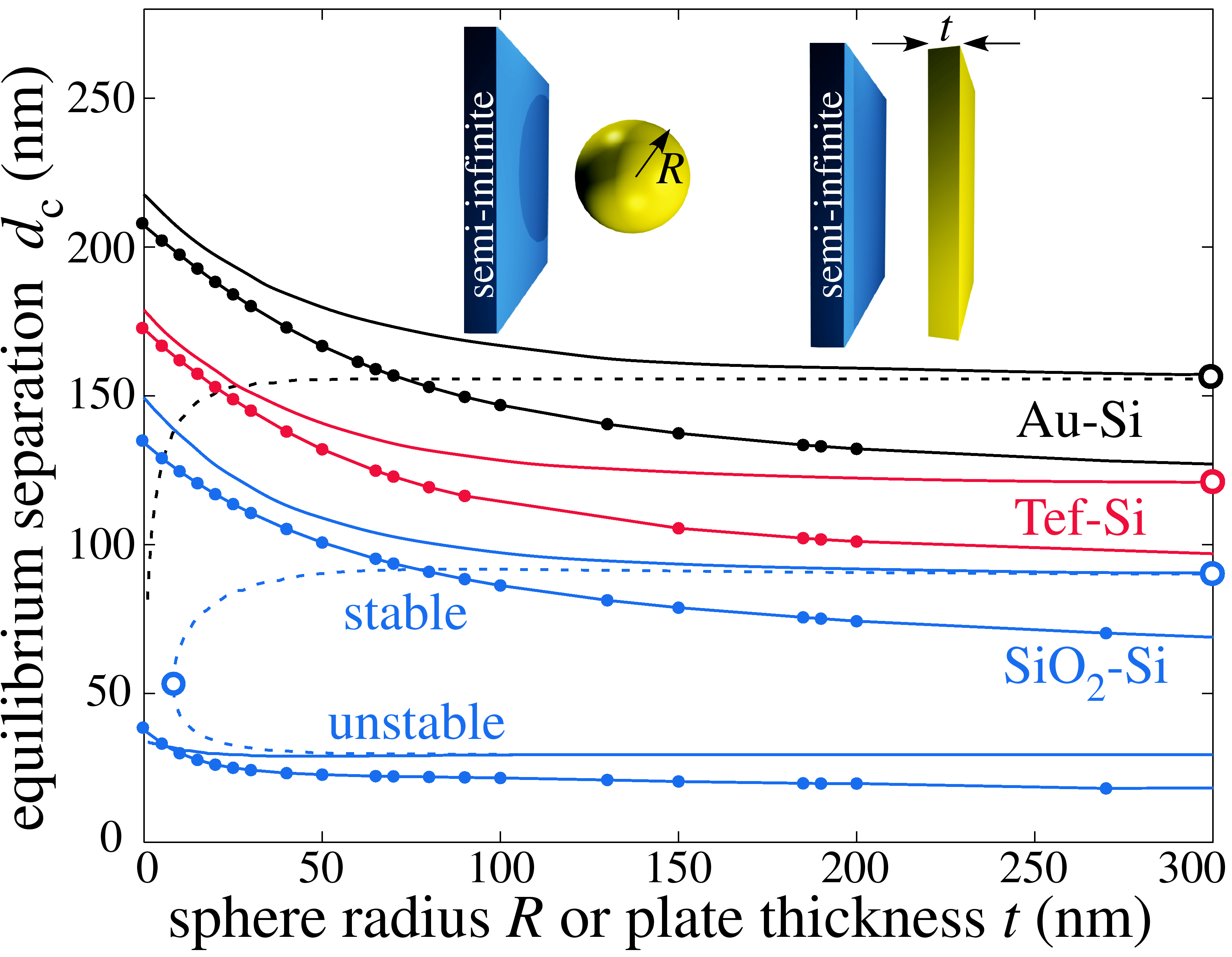}
\caption{Equilibrium separation $d_c$ vs. sphere-radius ($R$) or
  slab-thickness ($t$) for various slab--sphere (dots) and slab--slab
  (lines) configurations. Dashed lines correspond to slab--slab
  geometries with materials swapped (semi-infinite silicon slab).}
\label{fig:Rdc}
\end{figure}

The fact that $d_c$ is smaller in the slab--sphere case than for
semi-infinite slabs was initially unexpected since it contradicts the
intuition described above. However when we plot $d_c$ for the various
materials (solid dotted lines) as a function of $R \in (0, 350)$nm in
\figref{Rdc}, we indeed observe the expected behavior: as $R$
decreases, $d_c$ increases, asymptoting to a constant at $R=0$ that
corresponds to the Casimir--Polder force between a spherical
nanoparticle and a slab~\cite{Rahi09:PRD}.  Similar increases in $d_c$
as thickness $t$ is decreased are observed in \emph{some} of the
finite-slab geometries (solid lines) in \figref{Rdc}, at least in all
of the configurations where the thickness of the silicon is varied.
In the $t\to\infty$ limit for the slab--slab case, the semi-infinite
result is recovered.  For the slab--sphere case with $R\to\infty$, the
asymptotic $d_c$ occurs for a smaller separation than for
semi-infinite slabs: in this limit, where PFA is valid, the curvature
of the spheres yields an average separation that is larger than $d_c$.

An interesting question is to what extent one may tailor object
geometries in order to change the qualitative features of the force,
e.g. its sign. With this question in mind, the dashed lines in
\figref{Rdc} shows $d_c$ for slab--slab geometries in which the two
solid materials have been exchanged so that the seminfinite slab is
instead silicon.  Apparently, changing which slab is finite
qualitatively reverses the dependence on $d_c$ in some cases: for
Si--Au and Si--SiO$_2$ slabs, the equilibrium separation $d_c$
\emph{decreases} with decreasing thickness $t$, corresponding to an
\emph{increased} attractive force despite the fact that the attractive
contributions arise from small $\xi$ (which are intuitively cut off by
the finite thickness).  This reversal, however, does not happen in all
cases: it does not occur for the Si--Tef slab--slab geometry or for
\emph{any} of the slab--sphere geometries (and we do not plot $d_c$ in
these reversed cases because the results are very similar to the
original results).  Evidently, the finite \emph{lateral} size of the
spheres has a dramatic qualitative interaction with the material
dispersion. In future work, we plan to investigate this interesting
interplay between material and geometric dispersion, in addition to
the finite-size behavior exhibited by the finite
gold~\cite{LambrechtPi06} and SiO$_2$ slabs.

Another feature worth noting in \figref{Rdc} also stems from the
anomalous response of the Si--SiO$_2$ slab--slab geometry to changes
in the SiO$_2$ thickness $t$: the stable $d^{(s)}_c$ and unstable
$d^{(u)}_c$ equilibria ``collide'' at a critical radius $R_c$, below
which the force is purely attractive at all separations. Technically,
$R_c$ is the \emph{tipping point} of a saddle-point bifurcation and
the disappearance of both equilibria is an example of a fold
catastrophe~\cite{Arnold}.

\begin{figure}[t]
\includegraphics[width=1.0\columnwidth]{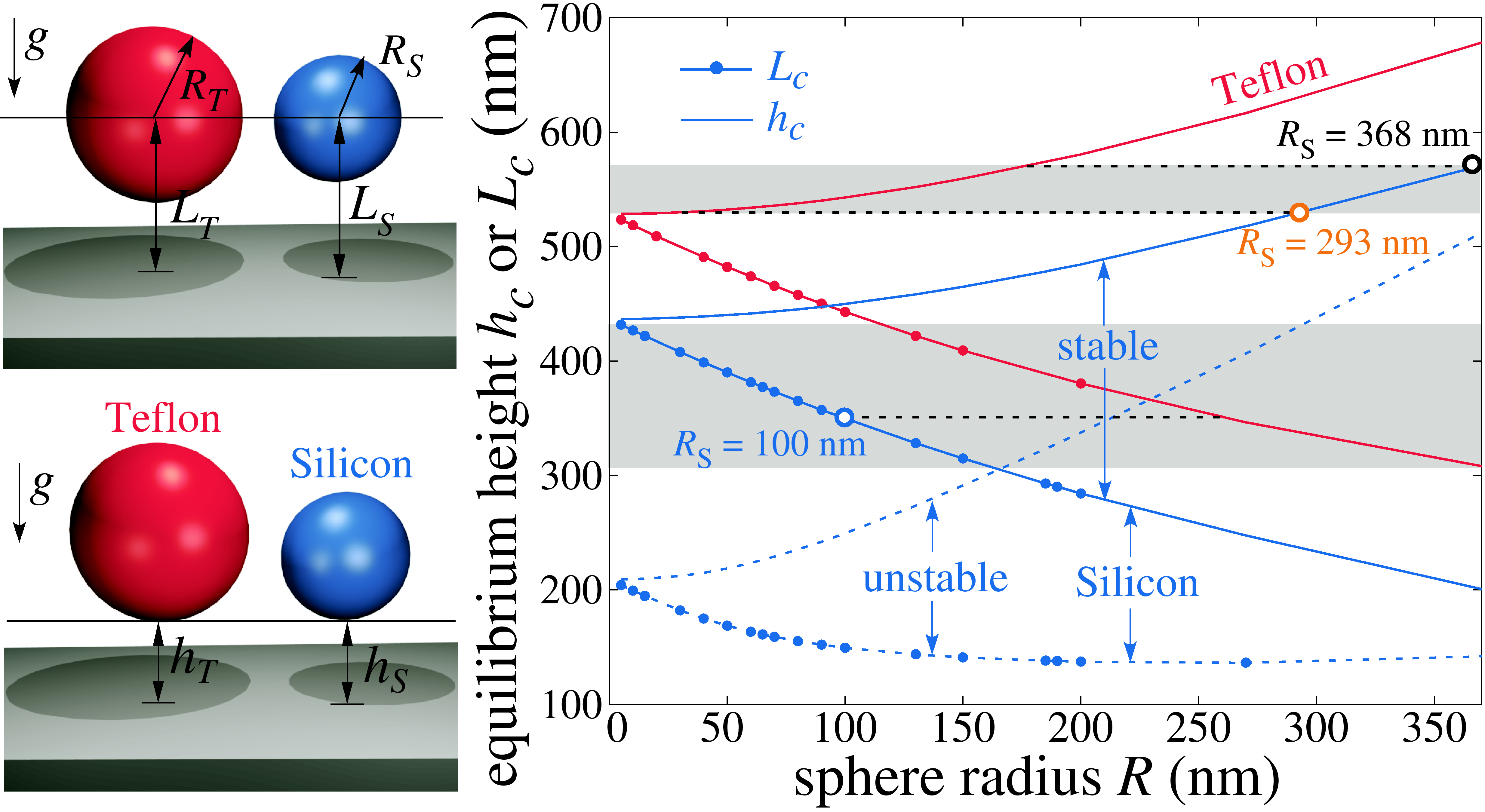}
\includegraphics[width=1.0\columnwidth]{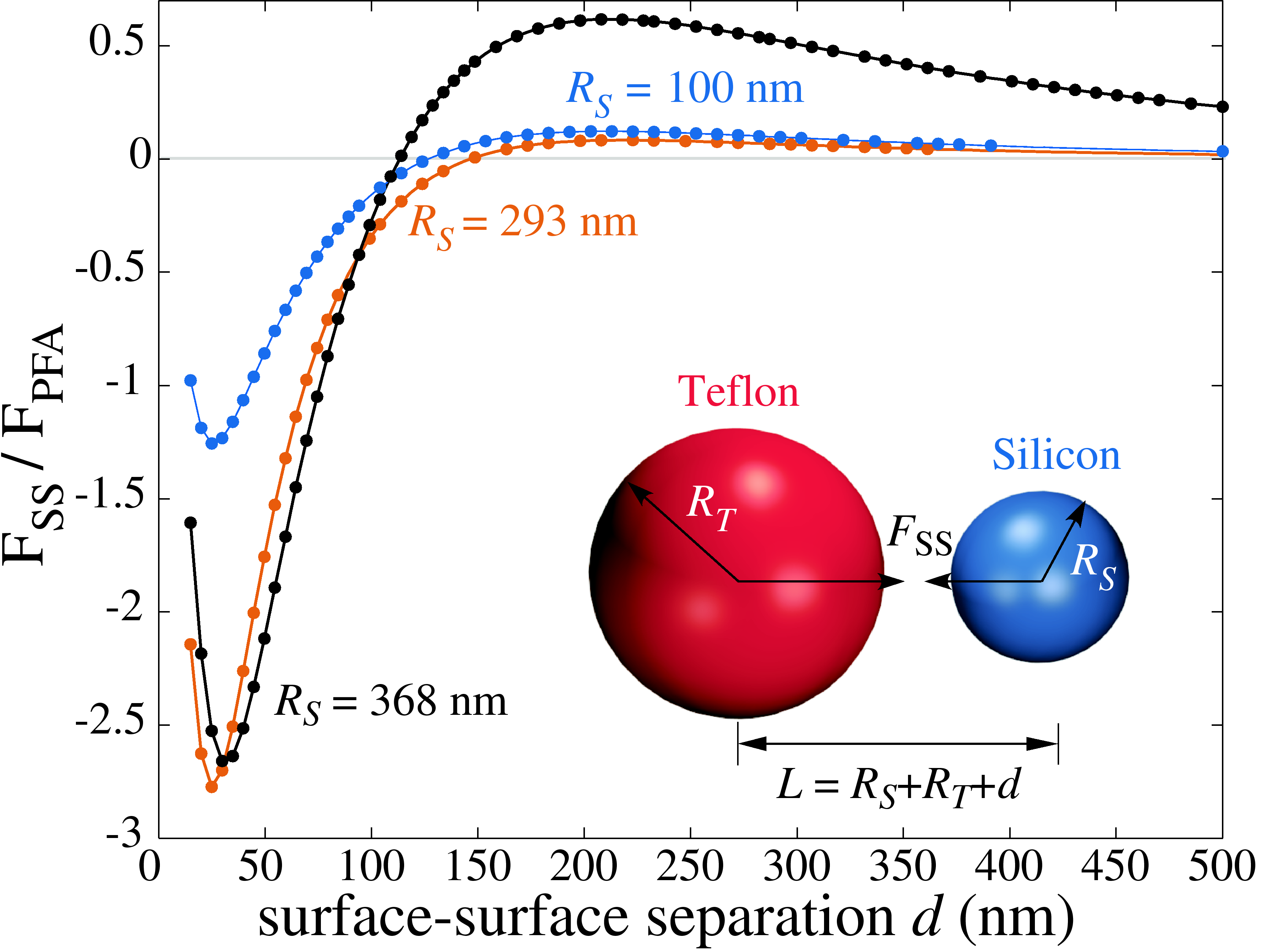}
\caption{(Top:) Plot of the stable equilibrium center--surface ($L_c$)
  and surface--surface ($h_c$) separation between either a teflon
  (yellow) or silicon (blue) sphere and a semi-infinite gold slab
  (depicted schematically on the left inset), as a function of sphere
  radius $R$. The black lines also show the presence of an unstable
  equilibrium in the silicon-sphere case. (Bottom:) Plot of the force
  $F_{SS}$ between two teflon/silicon spheres of radii $R_S$/$R_T$,
  showing the existence of a stable equilibrium.}
\label{fig:gravity}
\end{figure}

In what follows, we illustrate an interesting corollary of this type
of stability: the possibility of obtaining stable noncontact
configurations of compact objects at a nonzero separation. In
particular, we calculate the Casimir force in a nanoparticle-dicluster
system consisting of teflon and silicon spheres, of different radii
$R_T$ and $R_S$, respectively, immersed in ethanol. The force $F_{SS}$
between the spheres is plotted in \figref{gravity}(bottom) for
different sets of radii $R_S = \{99.69, 293.67, 368.39\}$nm and $R_T =
\{262.09, 32.67, 176.64\}$nm, respectively. This choice of the radii
was motivated by one possible experimental configuration, in which the
pair of spheres are also levitated above a planar slab: in this
geometry, discussed below, the sphere radii are chosen so that both
spheres are suspended at the same heigh above the slab.  With this
choice of materials, the spheres are again attractive at large
separations and repulsive for small separations, leading to a stable
(orientiation-independent) surface--surface separation in the
$100$--$150$nm range.

To make such a dicluster pair easier to observe in experiments, one
could simultaneously suspend them at a known distance above a planar
substrate, using the interplay between the repulsive Casimir force and
gravity to create stable levitation.  For simplicity, we investigate
this possibility within the additive approximation: the slab--sphere
and sphere--sphere interactions are considered independently. Because
Casimir forces are not additive, the presence of the slab will change
the stable separation of the spheres (and vice versa for the stable
height of the spheres). However, this approximation forms a useful
starting point for the design of such an experiment and should even be
accurate in the limit where the sphere diameter is much larger than
the stable surface--surface separations (here, most of the diameters
are at least twice the stable separations). \Figref{gravity}(top)
shows a plot of the equilibrium surface--surface ($h_c$) and
center-surface ($L_c = h_c + R$) separations of the teflon (yellow)
and silicon (blue) spheres suspended above a semi-infinite gold slab,
as a function of sphere radius $R$.  As shown by the figure,
decreasing $R$ acts to increase $h_c$ and decrease $L_c$. The decrease
in $h_c$ occurs much more rapidly than in \figref{Rdc} due to the fact
that, in addition to geometric dispersion, the force of gravity scales
as the mass $\rho V$ of a sphere (where the density $\rho \approx
\{2330, 2200, 789 \}$kg/m$^3$, for \{Si, teflon, ethanol\},
respectively). The center--center separation $L_c=h_c+R$ increases
with $R$ because $\partial h_c / \partial R > -1$.  In addition to a
stable equilibrium $h_c$, Au--Si exhibits an unstable equilibrium at
smaller $d_c$ due to the transition to an attractive Casimir force for
small separations; if the sphere were ever pushed below the unstable
equilibrium point, it would continue downward and adhere to the slab.
This would be a concern for experiments if fluctuations in the sphere
height could push it below the unstable equilibrium, but as seen from
\figref{gravity} the distance between the stable and unstable
equilibria is over $200$nm for $R < 50$nm nanoparticles.

The gray areas in \figref{gravity}(top) depict regions in which the
$L_c$ or $h_c$ of the two spheres can be made equal by an appropriate
choice of radii, as shown schematically in \figref{gravity}(top-left).
This determines the stable configuration of the two-sphere dicluster
when they are brought together above the surface; the three dashed
horizontal lines in \figref{gravity}(top) correspond to the radii used
for the force calculation in \figref{gravity}(bottom).


We are grateful to Jeremy Munday at Caltech and Jamal S. Rahi at MIT
for useful discussions. This work was supported by the Army Research
Office through the ISN under Contract No. W911NF-07-D-0004, and by US
DOE Grant No. DE-FG02-97ER25308 (AWR).



\end{document}